\documentclass[conference]{IEEEtran}
\usepackage{graphicx}
\usepackage{cite}
\usepackage{mathtools}
\usepackage{blindtext}
\usepackage{amsmath,amssymb,amsfonts}
\usepackage{algorithm}
\usepackage{algpseudocode}
\usepackage{setspace}
\usepackage{subcaption}	
\usepackage{textcomp}
\usepackage{balance}
\usepackage{xcolor}
\usepackage{multirow}
\usepackage{stfloats}
\usepackage{amsmath}
\usepackage{array}
\usepackage{url}
\usepackage{soul}
\usepackage{tabularx}
\usepackage{bm}
\usepackage{pifont}%
\renewcommand{\baselinestretch}{0.95}
\def\BibTeX{{\rm B\kern.05em{\sc i\kern-.025em b}\kern-.08em T\kern-.1667em\lower.7ex\hbox{E}\kern-.125emX}}

\begin{document}

\title{60~{GH}z Outdoor Propagation Measurements and Analysis Using Facebook Terragraph Radios}

\author{
\IEEEauthorblockN{Kairui Du\IEEEauthorrefmark{1}, Omkar Mujumdar\IEEEauthorrefmark{1}, Ozgur Ozdemir\IEEEauthorrefmark{1}, Ender Ozturk\IEEEauthorrefmark{1},\\ Ismail Guvenc\IEEEauthorrefmark{1}, Mihail L. Sichitiu\IEEEauthorrefmark{1}, Huaiyu Dai\IEEEauthorrefmark{1}, and Arupjyoti Bhuyan\IEEEauthorrefmark{2}}
\IEEEauthorblockA{\IEEEauthorrefmark{1}Department of Electrical and Computer Engineering, North Carolina State University, Raleigh, NC}
\IEEEauthorblockA{\IEEEauthorrefmark{2}INL Wireless Security Institute at Idaho National Laboratory}
Email: \{kdu, ommujumd, oozdemir, eozturk2, iguvenc, mlsichit, hdai\}@ncsu.edu, arupjyoti.bhuyan@inl.gov 
}

\maketitle

\begin{abstract}

The high attenuation of millimeter-wave~(mmWave) would significantly reduce the coverage areas, and hence it is critical to study the propagation characteristics of mmWave in multiple deployment scenarios. In this work, we investigated the propagation and scattering behavior of 60~GHz mmWave signals in outdoor environments at a travel distance of 98~m for an aerial link (rooftop to rooftop), and 147~m for a ground link (light-pole to light-pole). Measurements were carried out using Facebook Terragraph~(TG) radios. Results include received power, path loss, signal-to-noise ratio (SNR), and root mean square (RMS) delay spread for all beamforming directions supported by the antenna array. Strong line-of-sight (LOS) propagation exists in both links. We also observed rich multipath components (MPCs) due to edge scatterings in the aerial link, while only LOS and ground reflection MPCs in the other link.

\end{abstract}

\begin{IEEEkeywords}

60~GHz, AERPAW, channel sounding measurements, delay spread, drone, LOS, mmWave,  MPCs, path loss.

\end{IEEEkeywords}

\section{Introduction}

There have been extensive studies on mmWave propagation in recent years. Measurements at 28~GHz in New York~\cite{NYC} showed that penetration loss are affected by the distance, obstructions, and surrounding environments. Path loss data at 28~GHz was provided in another study~\cite{NYC2} based on measurements from 3 different transmitter~(Tx) locations and 75 receiver~(Rx) locations. In our recent 28~GHz measurements~\cite{library} at NC State University, path loss for the line-of-sight (LOS) scenarios was obtained to be very close to the free space path loss (FSPL) model. Models for large-scale path loss in both LOS and non-LOS (NLOS) scenarios were developed from measurements over distances ranging from 10~m to 50~m. The propagation and scattering behaviour of 28~GHz signals in indoor, outdoor, and indoor-to-outdoor environments at a regional airport were studied in~\cite{du2021airport}. Foliage and ground reflection measurements at 73~GHz~\cite{foliage} showed an average of 0.4~dB/m foliage attenuation, and reflection coefficients ranged from 0.02 to 0.34 for dirt and gravel ground.

Authors in~\cite{rappaport2015wideband} presented path loss models with directional and omnidirectional antennas based on over 15,000 measured power delay profiles (PDPs) at 28, 38, 60, and 73~GHz bands using wideband channel sounders. In~\cite{hosseini2020attenuation,du2021penetration}, authors focused on penetration loss of common constructional materials in both mmWave bands (28, 39, 73, 91~GHz) and sub-terahertz~(sub-THz) bands (120 and 144~GHz). Penetration loss was observed to increase with frequencies. Another study in~\cite{xing2019indoor} focused on indoor path loss, reflection, penetration, and scattering properties at mmWave (28~GHz and 73~GHz) and sub-THz (140~GHz) frequencies. Reflection loss was lower as frequency increased in a certain incident angle, while the penetration loss was also observed to increase with frequency.

All these studies indicate that mmWave features a high path loss and a high material attenuation. Deploying a number of base stations is preferred to overcome the propagation challenges and provide robust coverage and data rate performance at mmWave bands. Facebook recently introduced an affordable and easier to deploy solution to provide reliable and high data rate access in urban and suburban environments using wideband mmWave TG~\cite{Facebook} radios operating at 60~GHz~\cite{Nordrum2019TG,Shkel2021TG}, as shown in Fig.~\ref{TG_radio}(a). 
In~\cite{Aslam2020TG}, measurements in both  urban canyon and residential areas were performed with TG radios. Path loss in the urban canyon scenario was observed to be smaller when compared with the residential areas due to rich number of MPCs. Another paper~\cite{Beelde2020TG} presented the channel sounding results using TG radios in a bulk carrier vessel. The measured path loss was higher than the FSPL, and that high data rate can be fulfilled if the LOS path is not obstructed. Authors in~\cite{Tariq2020TG} measured path loss, received power, input and output SNR, and delay spread values for each specified beam combination with massive MIMO antenna arrays in both indoor and outdoor scenarios using TG radios.


In this paper, we conducted outdoor 60~GHz propagation measurements  at North Carolina State University, using Facebook TG sounders with phased-array antennas. Measurements were carried out between the equipment deployed at rooftops and light-poles. Path loss was observed to be 110.2 dB at a travel distance of 98 m for the aerial link, and was 117.05 dB at a separation of 147 m for the ground link. A peak SNR of 18 dB and 15 dB was achieved in the LOS region of the 2 links. Maximum RMS delay was 3.88 ns and 4.83 ns, respectively. We also observed strong LOS propagation and rich scatterings from the roof edge when the Tx and Rx were both placed on rooftop, while there are only LOS and ground reflection for the link at a lower height. 


\section{Measurement Setup and Configuration}
\label{sec_setup}

\begin{figure}[t]
 \centering
	\begin{subfigure}{0.24\textwidth}
	\centering
	\includegraphics[width=\textwidth]{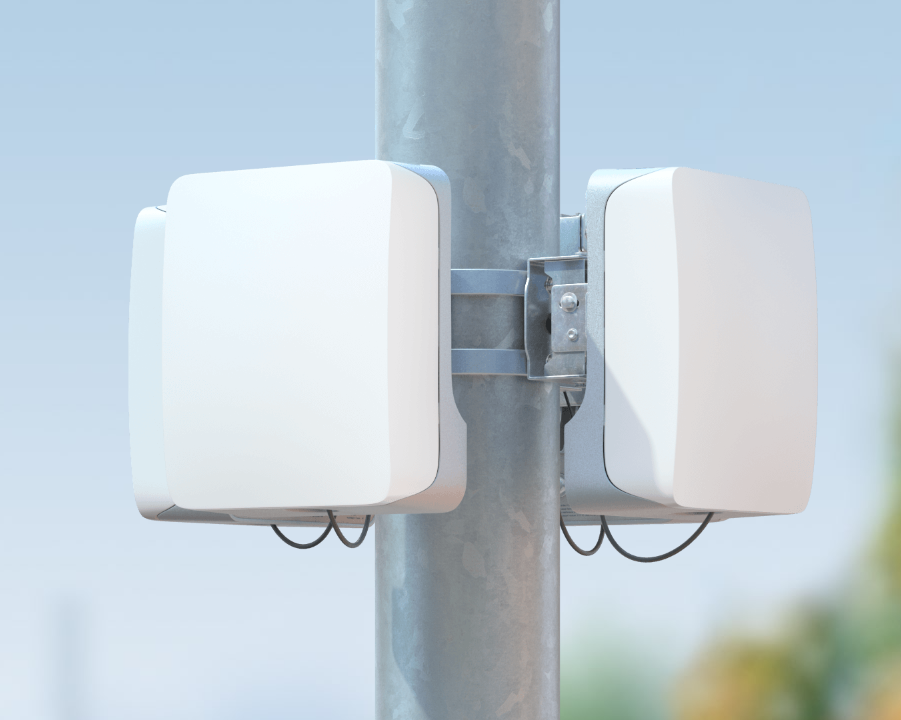} 
	\caption{}
    \end{subfigure}			
	\begin{subfigure}{0.24\textwidth}
	\centering
    \includegraphics[width=\textwidth]{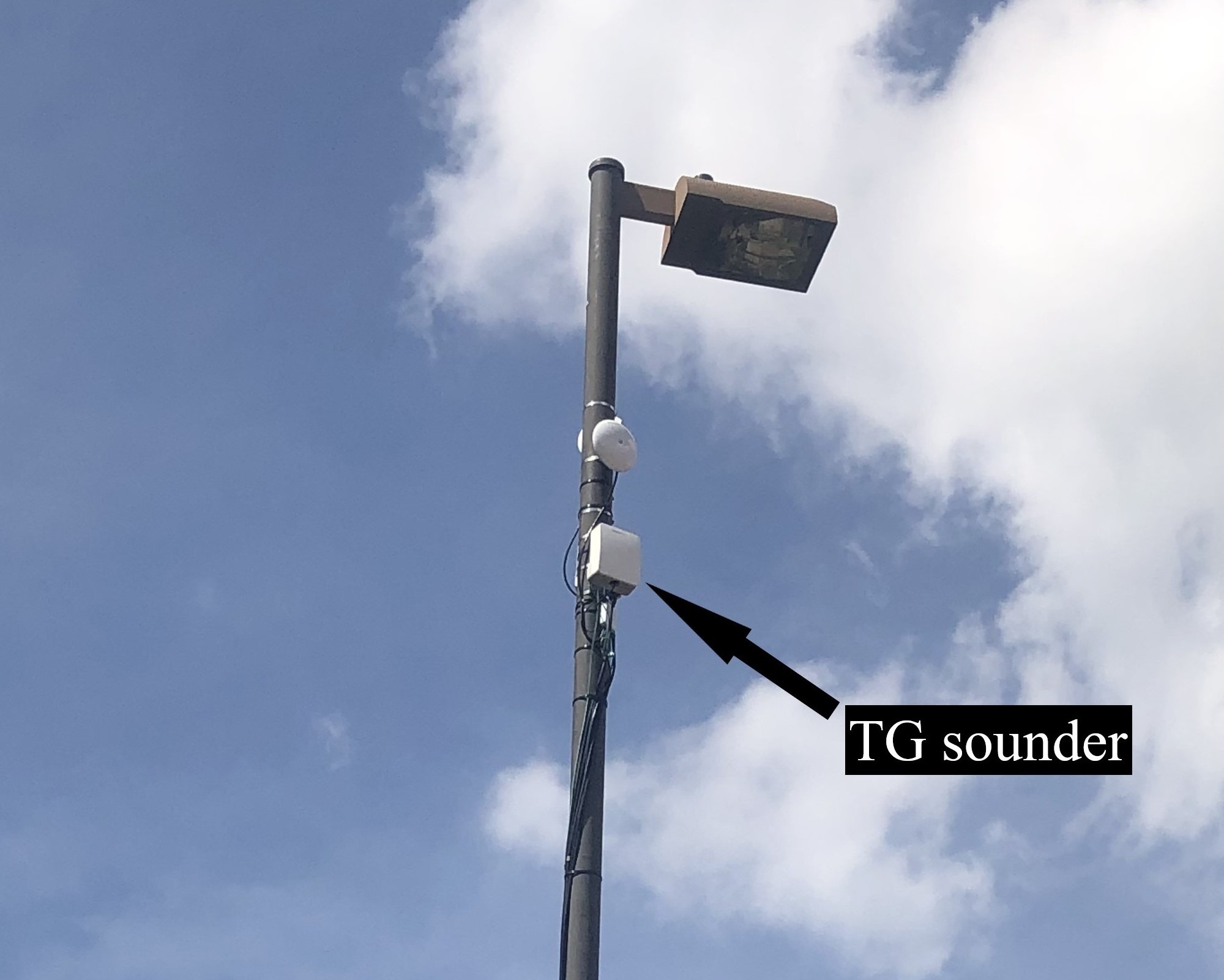}
	\caption{}
    \end{subfigure}
    \caption{(a) Facebook TG radios (b) TG sounder moundted on lamppost.}
    \label{TG_radio}
    \vspace{-3mm}
\end{figure}

The TG channel sounders are customized TG radios designed for measurement and modeling of 60~GHz channels. The sounder is capable of measuring the characteristics of the 60~GHz including the directional path loss and channel impulse response. The center frequency of the transmitted signal is 60.48~GHz with 2.16~GHz bandwidth. $36\times 8$ phased-array antennas were connected to both Tx and Rx. These antennas are comprised of 3 tiles, each tile is arranged as an array of 12 columns and 8 rows. All antennas in a given column are connected to the same phase shifter and can support a minimum of 2.8 degree half-power beamwidth (HPBW)~\cite{Shkel2021TG}. In our configuration, we used all 3 tiles for Tx and Rx antennas, and the array was steered in elevation plane. Both the Tx and the Rx antennas scanned an elevation angle of -45 degree to 45 degree, with 1.4 degree resolution. The HPBW of the array was 2.8 degree for both Tx and Rx arrays.


\begin{figure}[t]
 \centering
	\begin{subfigure}{0.24\textwidth}
	\centering
	\includegraphics[width=\textwidth]{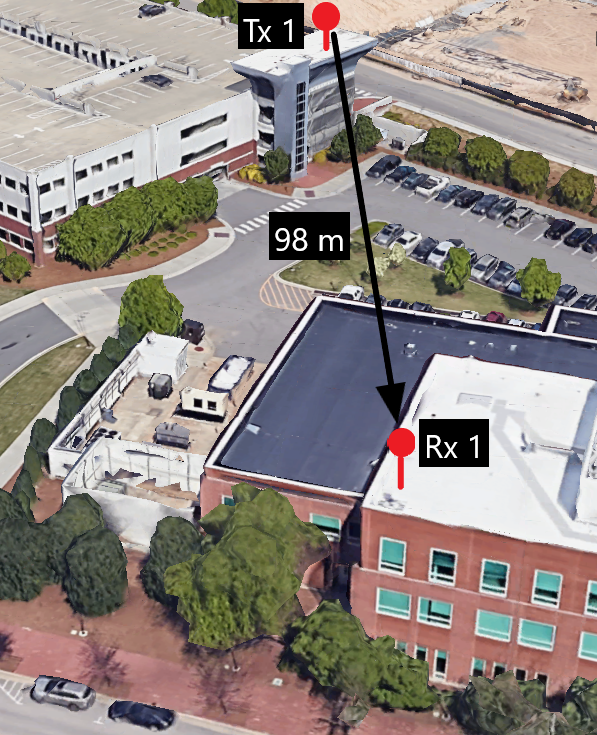} 
	\caption{}
    \end{subfigure}			
	\begin{subfigure}{0.24\textwidth}
	\centering
    \includegraphics[width=\textwidth]{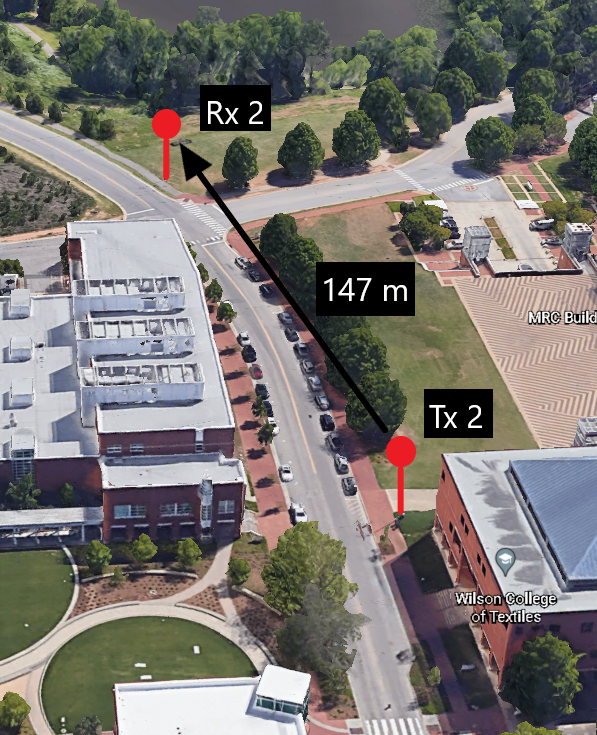}
	\caption{}
    \end{subfigure}
    \caption{The deployment of TG radios for (a) link 1: on the lampposts of rooftops with a separation of 98 m, Tx was 2 m above the rooftop of the parking deck and Rx was 2 m above a 3-floor building, (b) link 2: on the lampposts on grounds with a separation of 147~meters, the height was 10 m off the ground for both Tx and Rx.}
    \label{links_3D}
    \vspace{-3mm}
\end{figure}

TG sounders were mounted on the lampposts in outdoor environments, as shown in Fig.~\ref{TG_radio}(b). We captured the measurement data for 2 links, as shown in Fig.~\ref{links_3D}. The Tx and Rx were placed on the lampposts of the rooftop for link 1 with a separation of 98 m, and on the lampposts on the ground for link 2 with a separation of 147 m.  The Tx was 2 m off a parking deck rooftop and the Rx was 2 m off a 3-floor building for link 1. For link 2, both Tx and Rx were placed 10 m above the ground. The Tx and the Rx were manually aligned to each other for both links. The effective isotropic radiated power (EIRP) was 38.6~dBm for link 1 and it was 39.8~dBm for link~2.


\section{Results and Analysis}
\label{sec_results}

The summarized parameters and results for link~1 and link~2 are shown in Table~\ref{results}. Among all the Tx and the Rx antenna array combinations, link 1 had a highest received power of -71.69 dBm, a highest SNR of 18 dB, a lowest path loss of 110.2 dB, and a highest RMS delay spread of 3.88 ns. Link 2 had a highest received power of -77.24 dBm, a highest SNR of 15 dB, a lowest path loss of 117.05 dB, and a highest RMS delay spread of 4.83 ns.

\begin{table}[b]
\footnotesize
\centering
\caption{Measurement results for link 1 and 2.} 
\label{results}
\begin{tabular}
{|p{1.5in}|p{0.4in}|p{0.4in}|} \hline 
 \textbf{Parameters} & \textbf{Link 1} & \textbf{Link 2} \\ \hline
Distance (m)  &  97  &  147  \\ \hline 
EIRP (dBm)  &  38.6  &  39.8  \\ \hline 
Maximum received power (dBm)  &  -71.69  &  -77.24  \\ \hline 
Minimum path loss (dB)  &  110.2  &  117.05  \\ \hline 
Maximum SNR (dB)  &  18  &  15  \\ \hline 
Maximum RMS delay (ns)  &  3.88  &  4.83  \\ \hline 
\end{tabular}
\end{table}


\begin{figure}[t]
 \centering
	\begin{subfigure}{0.45\textwidth}
	\centering
	\includegraphics[width=\textwidth]{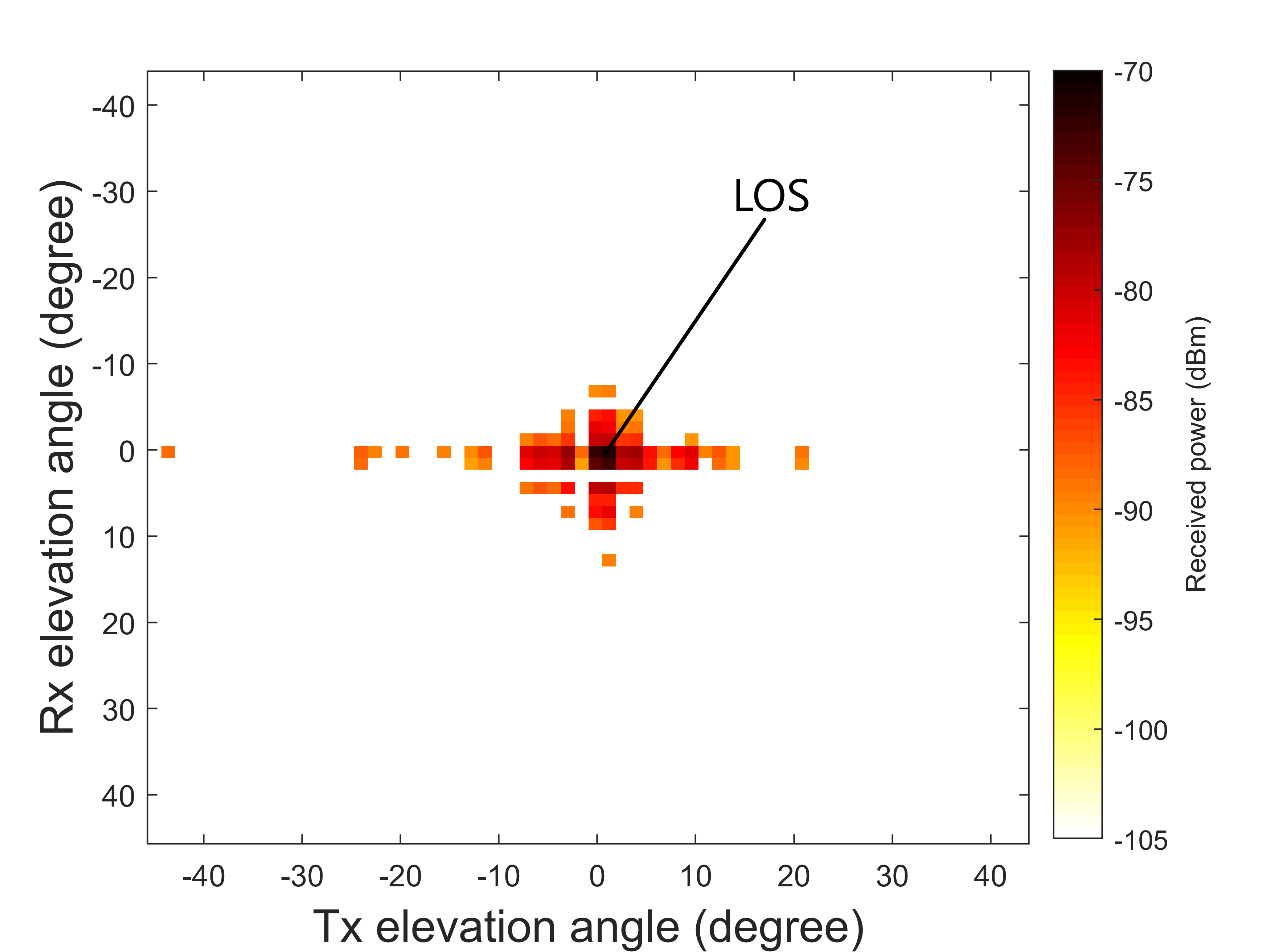} 
	\caption{}
    \end{subfigure}			
	\begin{subfigure}{0.45\textwidth}
	\centering
    \includegraphics[width=\textwidth]{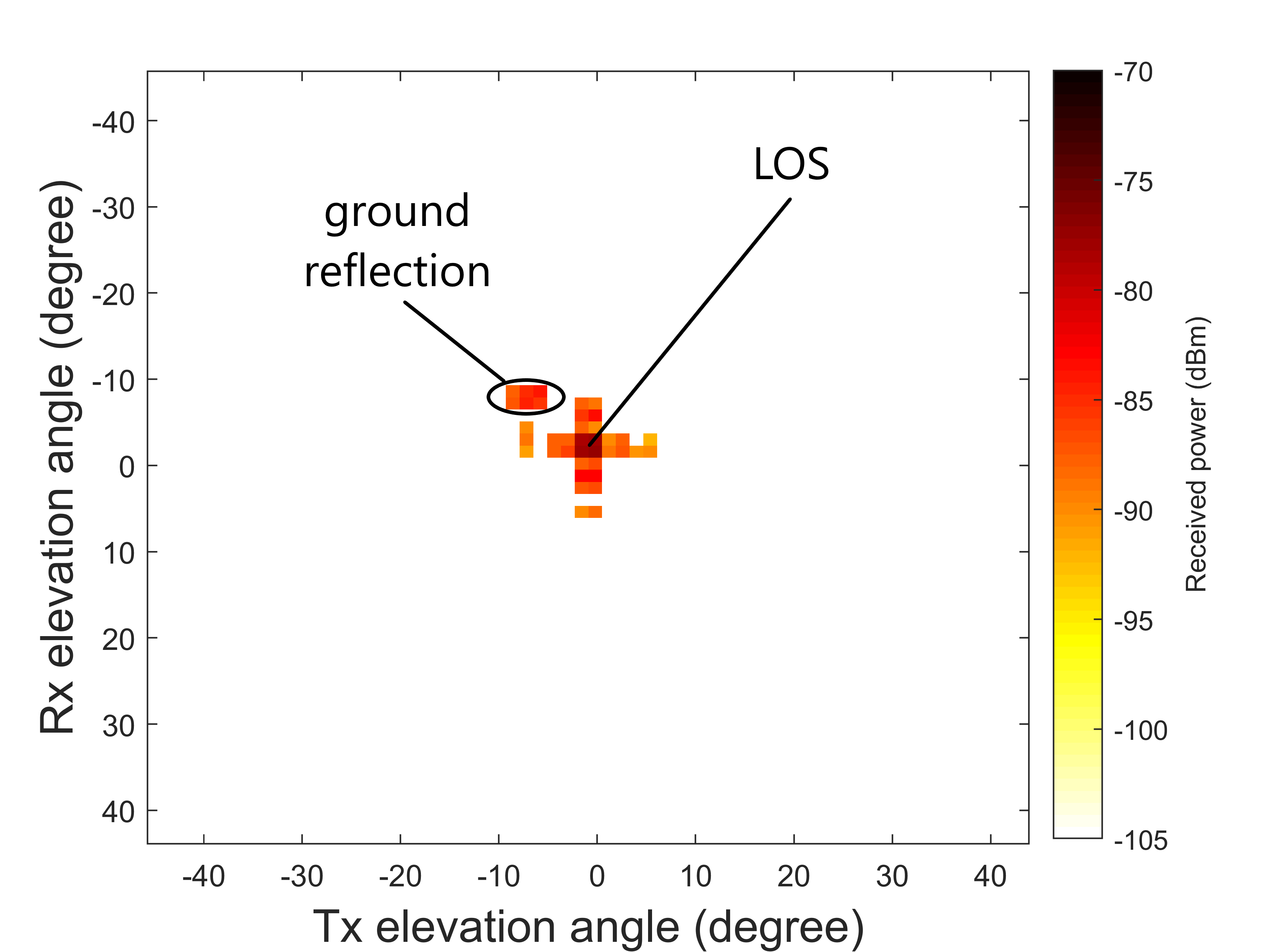}
	\caption{}
    \end{subfigure}
    \caption{Received power of (a) link 1, (b) link 2.}
    \label{reveived_power}
    \vspace{-3mm}
\end{figure}

The received power of the 2 links are shown in Fig.~\ref{reveived_power}. Each colored grid represents the received power at a given combination of Tx and Rx elevation angle. Both link observed a dominant LOS propagation. The maximum received power was -71.69~dBm for link 1, and was -77.24~dBm for link 2 when the Tx array and Rx array was directly pointing each other in the LOS region. As the misalignment between the Tx array and Rx array increased, the received power decreased and even not detectable. Link 1 has higher overall received power due to a closer distance. The wider coverage for link 1 was a sequence of edge scatterings from the building rooftop. Link 2, on the other hand, only observed ground reflection (Tx elevation angle at -8.6 degree and Rx elevation angle at -8.6 degree) when the Tx and Rx beams were not aligned. 


\begin{figure}[t]
 \centering
	\begin{subfigure}{0.45\textwidth}
	\centering
	\includegraphics[width=\textwidth]{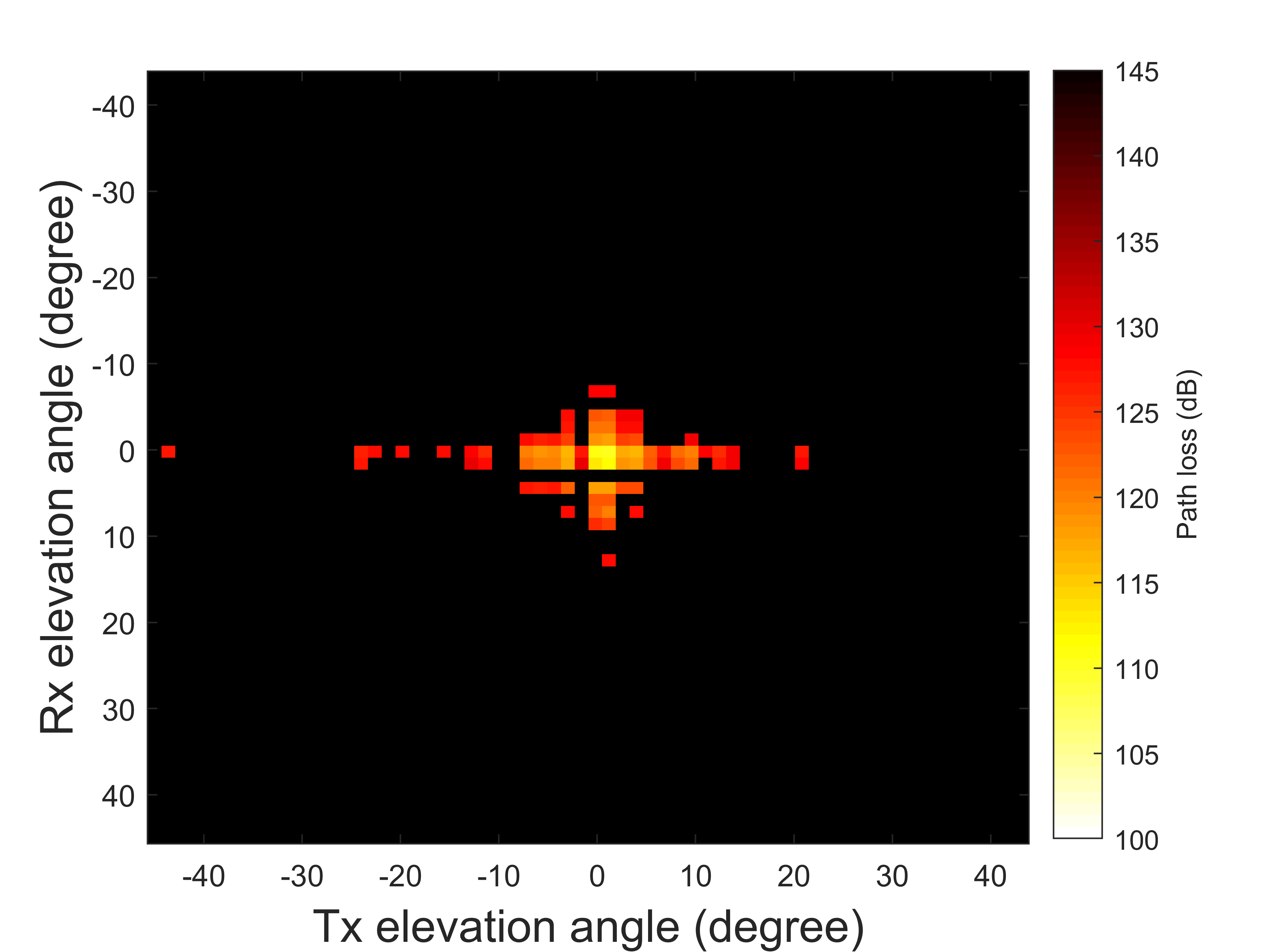} 
	\caption{}
    \end{subfigure}			
	\begin{subfigure}{0.45\textwidth}
	\centering
    \includegraphics[width=\textwidth]{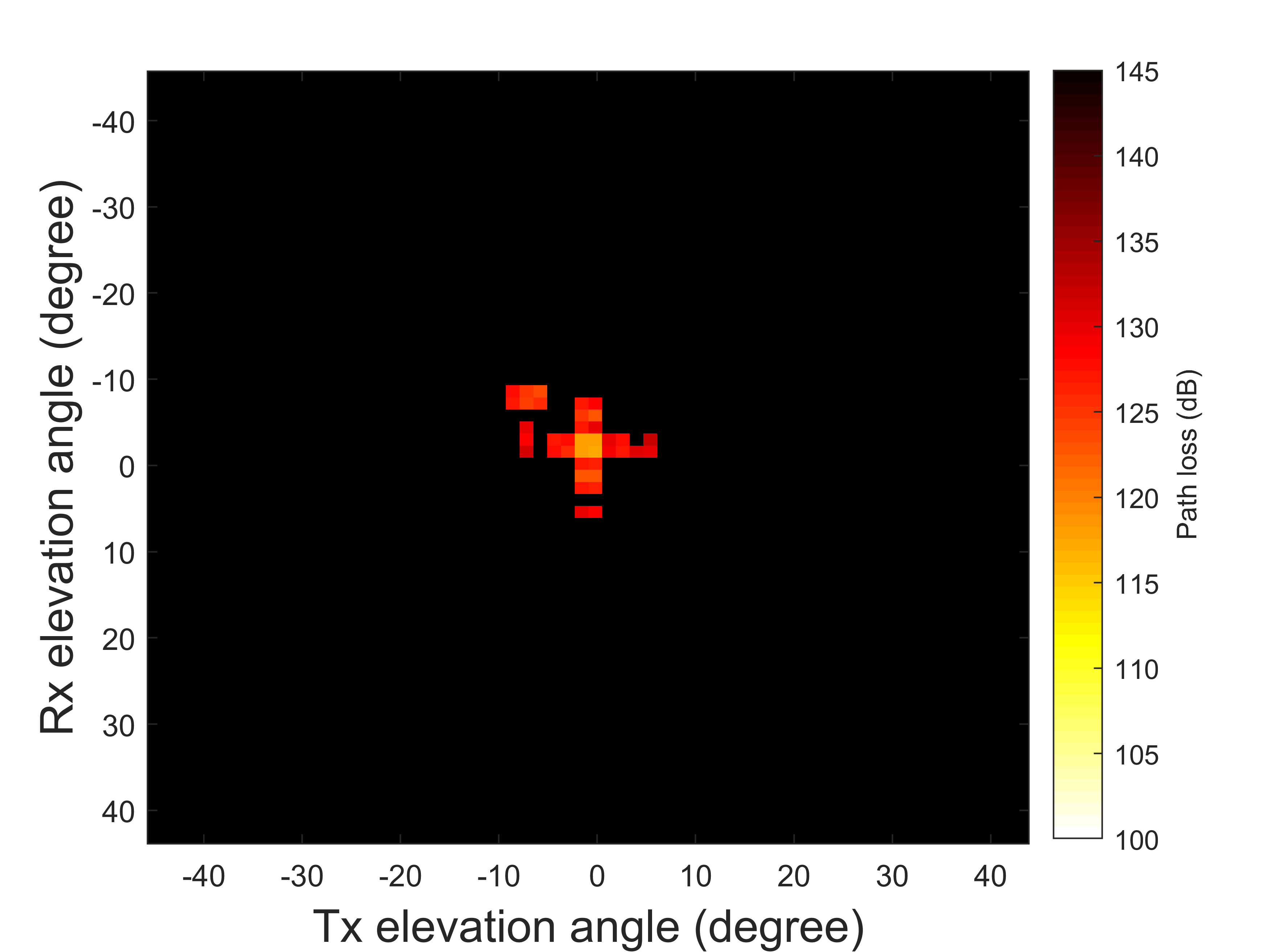}
	\caption{}
    \end{subfigure}
    \caption{Path loss of (a) link 1, (b) link 2.}
    \label{path_loss}
    \vspace{-3mm}
\end{figure}

The path loss at each beam combination are shown in Fig.~\ref{path_loss}. The minimum path loss was 110.2~dB for link 1, and was 117.05~dB for link 2 when the Tx array and Rx array was directly pointing each other in the LOS region. Path loss also increased as the misalignment between the Tx and Rx arrays increased. Overall, link 1 has a smaller path loss because the Tx and Rx separation was shorter.




\section{Conclusion}
\label{sec_conclusion}

In this work, we performed channel measurements of 60~GHz mmWave for an aerial link with the Tx and the Rx placed 2 m off the rooftops, and a ground link with the Tx and the Rx both at a height of 10 m. Results show a path loss of 110.2 dB, an SNR of 18 dB, and a maximum RMS delay of 3.88 ns at a separation of 98 m for the aerial link, and that of 117.05 dB, 15 dB, and 4.83 ns for the ground link with a separation of 147 m. Strong LOS connectivity exists in both links, while the aerial link has richer number of scatterings from the roof edge, and the ground link only has a ground reflection. If the paper is accepted, we will include additional results based on theoretical path loss models and measurements/discussions for an air-to-ground link (rooftop to light-pole, between RX1 and TX2) that was not operational initially. These links can be representative of propagation environments for low-altitude drones. Our future work includes opening the Facebook TG platform for external user experimentation under the NSF AERPAW project.

\section*{Acknowledgement}

This work was supported in part by NSF CNS-1939334, and by INL Laboratory Directed Research and Development (LDRD) Program under DOE Idaho Operations Office Contract DEAC07-05ID14517.

\bibliographystyle{IEEEtran}
\bibliography{IEEEabrv,reference}

\end{document}